\documentclass[twocolumn,english,aps,prl,showpacs]{revtex4}
\usepackage[T1]{fontenc}
\usepackage[latin9]{inputenc}
\usepackage{amsmath}
\usepackage{amssymb}
\usepackage{graphicx}
\usepackage{esint}
\usepackage{bm}
\usepackage{babel}

\makeatletter
\newcommand{\lyxdot}{.}
\@ifundefined{textcolor}{}
{
 \definecolor{BLACK}{gray}{0}
 \definecolor{WHITE}{gray}{1}
 \definecolor{RED}{rgb}{1,0,0}
 \definecolor{GREEN}{rgb}{0,1,0}
 \definecolor{BLUE}{rgb}{0,0,1}
 \definecolor{CYAN}{cmyk}{1,0,0,0}
 \definecolor{MAGENTA}{cmyk}{0,1,0,0}
 \definecolor{YELLOW}{cmyk}{0,0,1,0}
}
\makeatother

\input{tcilatex}

\begin{document}

\title{Why asymmetric interparticle interaction can result in convergent
heat conductivity }
\author{Shunda Chen}
\author{Yong Zhang}
\author{Jiao Wang}
\author{Hong Zhao}
\email{zhaoh@xmu.edu.cn}
\affiliation{Department of Physics and Institute of Theoretical Physics and Astrophysics,
Xiamen University, Xiamen 361005, Fujian, China}

\begin{abstract}
We show that the asymmetric inter-particle interactions may induce rapid
decay of heat current autocorrelation in one-dimensional momentum conserving
lattices. When the asymmetry degree and the temperature are appropriate, the
decay is sufficient rapid for resulting a convergence conductivity
practically. To understand the underlying mechanism, we further studied the
relaxation behavior of the hydrodynamic modes. It is shown that for lattice
with symmetric potential, the heat mode relaxs in the superdiffusive manner,
while in the case of asymmetric potential, the heat mode may relax in the
normal manner.
\end{abstract}

\date{\today }
\maketitle

In recent decades,the heat transport properties of low-dimensional lattice
models have been a particularly interesting issue. Based on intensive
theoretical analysis \cite{Pros,Nara,Mai,JSW,Gray,Del,Beijeren12} and
numerical studies (see for example \cite%
{Lepri97,Zhao98,Lebowitz,Lepri,Dharrev} and references therein), for 1D
momentum conserving fluids and lattices, at present it is generally accepted
that the heat current autocorrelation should decay in power-law manner. An
important consequence of the power-law decay is the divergence of the heat
conductivity following the Green-Kubo formula. Recently, we found that
lattice models with asymmetric inter-particle interactions may induce the
rapid decay of the current-current correlation function \cite%
{arxivV1,arxivV2,arxivV3} and results in size-independent conductivity \cite%
{Zhong,CPB}. In fact, lattice models with asymmetric interactions have been
studied in the literature both analytically \cite{Pros,Gray,Del,Beijeren12}
and numerically \cite{Del,Gray,Pros,FPUAB}. In particular, the hydrodynamics
analysis based on the Burgers equation suggests a divergent conductivity
even for systems with asymmetric interactions, which agrees with the result
based on the Zwangzig-Mori equation and the self-consistent mode coupling
theory \cite{Del,Beijeren12}. Recently, several works appear in arXiv \cite%
{arxivSavin,arxivDhar,arxivWL}, declared a partial or even complete
disagreement to our conclusion.

This paper present further explaination to our point of view. We explain why
we can conclude that the ripid decay of the current-current correlation may
result a convergent conductivity practically. To understand the mechanism
and aviod the non-physical extrapolation we emphasize that one should
distinguish the heat and energy. The heat mode and sound mode may behave
differentlly between lattice models with symmetric and that with asymmetric
potentials.

However, to have a solid basis to discuss or dispute, we would like to
re-declare several key points of our views before present the new result.

We declared that the asymmetry interactions may result size-independent heat
conductivity for one-D lattices. Not must be always. The lattice models we
have studied include the  FPU-$\alpha \beta $ model, the L-J lattice models,
and several other  asymmetry interaction models. \cite%
{arxivV1,arxivV2,arxivV3,Zhong,CPB} For these models, we have shown that
with \textbf{proper} degree of asymmetric potential, the current-current
correlation function decays in a manner faster than the power law, and the
heat conductivity becomes size-independent when the sytstem size is big
enough. We emphsized the temperature should not too high and the potential
parameters should be proper in the FPU-$\alpha \beta $ model to guaratee the
proper degree of the asymmetry. We agree that in the higher temperature case
the heat conductivity may \textbf{still divergent} as in the case of the FPU-%
$\beta $ model.

We declared that the conclusion is for \textbf{lattice} model, not for
liquid model. We agree that for the gas model with asymmetry interaction,
the conductivity diverges with the system size.

We pointed out that the so-called low-temperature region is corresponding to
the room temperature. So our finding has practically importance.

We first explain why practically the rapid decay of the current correlation
function can induce a size-independent conductivity. Here we take only the
Fermi-Pasta-Ulam-$\alpha$-$\beta$ (FPU-$\alpha$-$\beta$) model to be an
example, though we have also studied a lot of asymmetry-potential lattice
models such as the Lennard-Jones (L-J) model \cite%
{arxivV1,arxivV2,arxivV3,Zhong,CPB,unpub}. The model is defined by the
Hamiltonian 
\begin{equation}
H=\sum_{i}\frac{p_{i}^{2}}{2}+V(x_{i}-x_{i-1}-1),
\end{equation}
where $p_{i}$ and $x_{i}$ are the momentum and position of the $i$th
particle, respectively, and $V$ is the potential between two neighboring
particles. The component particles are assumed to be identical and have unit
mass, and the lattice constant is set to be unity so that the system length $%
L$ equals the particle number $N$. The inter-particle interactions in the
FPU-$\alpha$-$\beta$ model are given by 
\begin{equation}
V(x)=\frac{1}{2}x^{2}-\frac{\alpha}{3}x^{3}+\frac{1}{4}x^{4},
\end{equation}
where the parameter $\alpha$ controls the degree of asymmetry as illustrated
in Fig. 1(a). For $\alpha=0$ the system reduces to the Fermi-Pasta-Ulam-$%
\beta$ (FPU-$\beta$) model with symmetric potential only. To well reveal the
effects of the asymmetry in this model, in our simulations the average
energy per particle, denoted by $\varepsilon$, is fixed to be $%
\varepsilon=0.1$ such that the averaged potential energy per particle is
about $0.05$.

The energy current \cite{Dhar} is defined as $J_{q}\equiv\sum_{i}\dot{x}_{i}%
\frac{\partial V(x_{i},x_{i+1})}{\partial x_{i}}.$ For a lattice the energy
current is equal to the heat current because there is no residual global
velocity \cite{Lepri}. To numerically measure the current in equilibrium
state, the system is first evolved from an appropriately assigned random
initial condition for a long enough time ($>10^{8}$) to ensure that it has
relaxed to equilibrium state; then the current at ensuing times is measured.
The periodic boundary condition is applied in the calculations, and the
total momentum is set to be zero.

In Fig. 1 we show the autocorrelation function of the energy current $%
C(t)\equiv \frac{\langle J(t)J(0)\rangle }{\langle J(0)J(0)\rangle }$ for
high and low temperatures respectively. The results are presented in log-log
scale. In generating Fig. 1, the system size is fixed to be $N=2048.$ It can
be seen that in the high temperature cases the correlation function decays
in power-law $C_{qq}(t)\sim t^{-\gamma },$ which agrees qualitatively with
previous studies and theoretical predictions \cite%
{Nara,Dhar,Del,Beijeren12,arXivSpohn}. Therefore in high-temperature case
the long-time-tail prediction applies. However, at low-temperature case, we
can see that the decay becomes faster than the power-law manner (see also
Fig. 2), which can be roughly regarded to be exponential.

\begin{figure}[tbp]
\vskip-0.1cm \includegraphics[width=1.05\columnwidth]{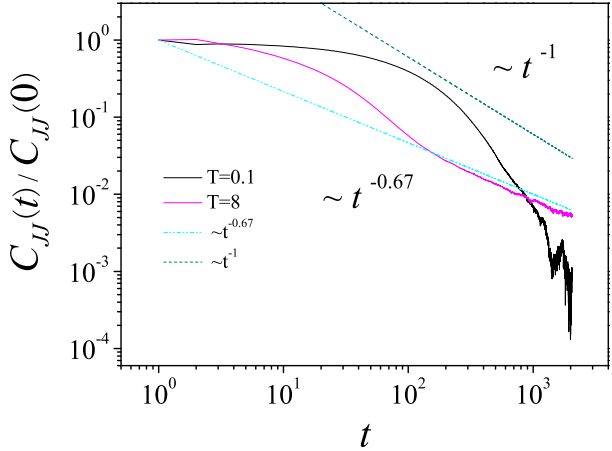} \vskip%
-0.4cm 
\caption{(Color online) Temperature dependence of the decay behavior of
current correlation function for the FPU-$\protect\alpha$-$\protect\beta$
lattice model with $\protect\alpha=2$, $\protect\beta=1$. }
\end{figure}

\begin{figure}[tbp]
\vskip-0.1cm \includegraphics[width=1.05\columnwidth]{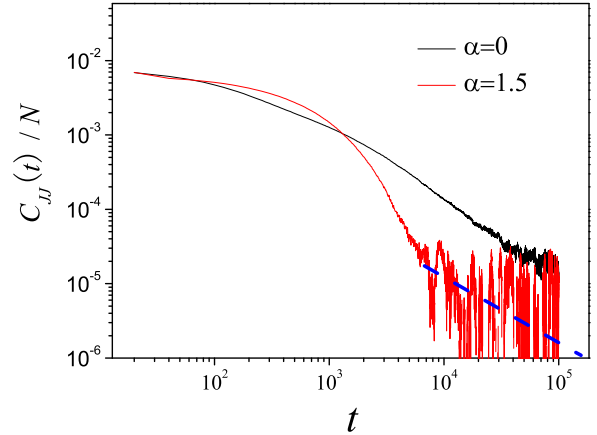} %
\vskip-0.4cm 
\caption{(Color online) Plots of the current correlation function for the
FPU-$\protect\alpha$-$\protect\beta$ lattice model ($\protect\alpha=0$
corresponds to FPU-$\protect\beta$ model).(T=0.1). }
\end{figure}

\begin{figure}[tbp]
\vskip-0.1cm \includegraphics[width=1.05\columnwidth]{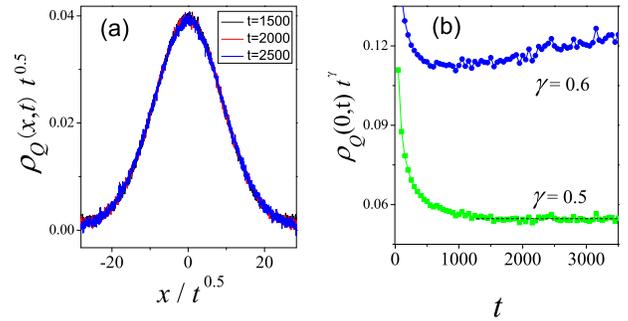} \vskip%
-0.4cm 
\caption{(Color online) Plots of the heat diffusion the FPU-$\protect\alpha$-%
$\protect\beta$ lattice model with $\protect\alpha=2$,$\protect\beta=1$. $%
\protect\gamma=0.5$ indicates normal heat diffusion.}
\end{figure}

Some researchers argue that it could not conclude that the correlation
function may still become as power-law decay when the evolution time is long
enough. Yes, with numerical simulations we can not exclude such a
possibility. But, what can be concluded is that such a long-period of rapid
decay as shown in the figure can already insure a size-independ conductivity
practically. To explain this, it is better to divide the Green-Kube formula
into two parts,

$\kappa=\lim\limits _{\tau\rightarrow\infty}\lim\limits _{L\rightarrow\infty}%
\frac{1}{2k_{B}T}\int_{0}^{\tau}C(t)dt\symbol{126}\frac{1}{2k_{B}T}%
[\int_{0}^{\tau_{e}}C(t)dt+$ $\int_{\tau_{e}}^{\tau_{tail}}C(t)dt]$

where $\tau_{e}$ represents the exponential-decay time period and $%
(\tau_{e},\tau_{tail})$ represents the tail. In the case of $T=0.1$, it has $%
\tau_{e}\symbol{126}2000$. Therefore, it has $\int_{0}^{\tau_{e}}C(t)dt%
\symbol{126}123$ by direct integral. The second part, even assume it decays
as the power-law of $\symbol{126}t^{-0.67}$, it contribute a sum of $%
\int_{\tau_{e}}^{\tau_{tail}}C(t)dt\symbol{126}10^{-4}t^{0.33}|_{\tau_{e}}^{%
\tau_{tail}}$. In this case, with even $\tau_{tail}=10^{12}$ it has $%
\int_{\tau_{e}}^{\tau_{tail}}C(t)dt\symbol{126}1$. To correlate the result
of the Green-Kube formula to that of the dirrect nonequilibrium simulation,
it is usually suggested \cite{ProsenChaos05} that the integral time should
be truncated at $\tau_{tail}=L/v$, where $v$ is the sound speed of the
system. Therefore, even the tail is power-law, the conductivity may turn to
power-law divergence till at least $L>$ $10^{12}$, which is about $100$
meter length and thus physically meanless.

The fundamental machnism should be understood by studying the relax behavior
of hydrodynamic modes \cite{ZhaoDiffusion,ChenDiffusion,ScienceChina}.
According to the hydrodynamical theory, fluctuations of a physical quantity
will relax as a superposition of the hydrodynamical modes of heat and sound.
To calculate the modes, one can apply the method described in \cite%
{ChenDiffusion,ScienceChina}. In such a way, we obtain the scaling exponent $%
\gamma$ of the Prähofer-spohn scaling function for several one-dimension
systems.

In Fig. 3 (a)-(b), we show the relax behavior of the heat mode in the case
of the FPU-$\alpha$-$\beta$ model as an example. These plots indicate that
the heat mode relax in the normal manner with $\gamma=0.5$.

This is the fundamental reason why the heat current in the FPU-$\alpha$-$%
\beta$ model may still obey the Foure law at low-temperature region. Our
studies remind us to distinguish the heat part and the sound part to be
transported. In the lattice models with proper degree and asymmetry, the
heat part relax in the normal manner, while the sound may still be
transported to infinite. The energy current carried by the sound mode thus
may decay in the power-law manner. In this case, the heat in a heat bath is
transported following the Foure heat conduction law. In real material, one
would find a normal heat conduct behavior. Meanwhile, there are a small part
of sound energy in the heat bath, no matter how small it is, it will
contribute a power-law decay energy current. When applying the Green-Kube
formula, it will result the divergence of the integral. This is a
non-physical effect and should be avioded, instead to declare a abnormal
heat conduct behavior for this kind of systems.

Very useful discussions with H. Spohn and A. Dhar are gratefully
acknowledged. This work is supported by the NNSF (Grants No. 10925525, No.
11275159, and No. 10805036) of China.


\begin{thebibliography}{99}
\bibitem{Pros} T. Prosen and D. K. Campbell, Phys. Rev. Lett. \textbf{84},
2857 (2000).

\bibitem{Nara} O. Narayan and S. Ramaswamy, Phys. Rev. Lett. \textbf{89},
200601 (2002).

\bibitem{Mai} T. Mai and O. Narayan, Phys. Rev. E \textbf{73}, 061202 (2006).

\bibitem{JSW} J. S. Wang, B. Li, Phys. Rev. Lett. \textbf{92}, 074302 (2004).

\bibitem{Beijeren12} H. van Beijeren, Phys. Rev. Lett. \textbf{108}, 180601
(2012).

\bibitem{Del} L. Delfini, S. Lepri, R. Livi, and A. Politi, Phys. Rev. E 
\textbf{73}, 060201 (2006); J. Stat. Mech. P02007 (2007).

\bibitem{Gray} G. R. Lee-Dadswell, B. G. Nickel, and C. G. Gray, Phys. Rev.
E \textbf{72}, 031202 (2005); J. Stat. Phys. \textbf{132}, 1 (2008).

\bibitem{Lepri97} S. Lepri, R. Livi, and A. Politi, Phys. Rev. Lett. 78,
1896 (1997) ; Europhys. Lett. 43, 271 (1998).

\bibitem{Zhao98} B. Hu, B. Li, and H. Zhao, Phys. Rev. E 57, 2992 (1998).

\bibitem{Lebowitz} F. Bonetto, J. L. Lebowitz, L. Rey-Bellet, \textit{%
Fourie's Law: A Chanllenge to Theorists}, Mathematical Physics 2000
(Imperial College Press, London, 2000).

\bibitem{Lepri} S. Lepri, R. Livi, A. Politi, Physics Reports \textbf{377},
1 (2003).

\bibitem{Dharrev} A. Dhar, Adv. Phys. \textbf{57}, 457 (2008).

\bibitem{arxivV1} S. Chen, Y. Zhang, J. Wang, and H. Zhao, ArXiv:1204.5933v1.

\bibitem{arxivV2} S. Chen, Y. Zhang, J. Wang, and H. Zhao, ArXiv:1204.5933v2.

\bibitem{arxivV3} S. Chen, Y. Zhang, J. Wang, and H. Zhao, ArXiv:1204.5933v3.

\bibitem{Zhong} Y. Zhong, Y. Zhang, J. Wang, and H. Zhao, Phys. Rev. E 
\textbf{85}, 060102(R) (2012).

\bibitem{CPB} Y. Zhong, Y. Zhang, J. Wang, and H. Zhao, Chin. Phys. B 22,
070505 (2013).

\bibitem{FPUAB} L. Wang and T. Wang, Europhys. Lett. \textbf{93}, 54002
(2011).

\bibitem{arxivSavin} A. V. Savin, Yuriy A. Kosevich, arXiv:1307.4725. 

\bibitem{arxivDhar} S. G. Das, A. Dhar, O. Narayan, arXiv:1308.5475.

\bibitem{arxivWL} L. Wang, B. Hu, B. Li, arXiv:1308.6061.

\bibitem{unpub} S. Chen, Y. Zhang, J. Wang, and H. Zhao, (unpublished).

\bibitem{Dhar} T. Mai, A. Dhar, and O. Narayan, Phys. Rev. Lett. \textbf{98}%
, 184301 (2007).

\bibitem{arXivSpohn} H. Spohn, arXiv:1305.6412.

\bibitem{ProsenChaos05} T. Prosen and D. K.Campbell, 15,015117 (2005).

\bibitem{ZhaoDiffusion} H. Zhao, Phys. Rev. Lett. 96, 140602 (2006).

\bibitem{ChenDiffusion} S. Chen, Y. Zhang, J. Wang, and H. Zhao, Phys. Rev.
E 87, 032153 (2013).

\bibitem{ScienceChina} S. Chen, Y. Zhang, J. Wang, and H. Zhao, Sci
China-Phys Mech Astron, 56: 1466\textendash{}1471 (2013). doi:
10.1007/s11433-013-5163-9.
\end{thebibliography}
\end{document}